\documentclass[11pt]{article}

\usepackage[utf8]{inputenc}
\usepackage[T1]{fontenc}
\usepackage[a4paper,margin=1in]{geometry}
\usepackage{amsmath,amssymb,bm,mathtools}
\usepackage{booktabs,graphicx,array}
\usepackage[authoryear,round]{natbib}
\usepackage[hidelinks]{hyperref}

\newcommand{\cl}{\mathrm{cl}}
\newcommand{\Kn}{\mathit{Kn}}
\newcommand{\dd}{\mathrm{d}}

\newcommand{\Aten}{\boldsymbol{A}}

\newcommand{\qv}{\boldsymbol{q}}
\newcommand{\cv}{\boldsymbol{c}}
\newcommand{\uv}{\boldsymbol{u}}
\newcommand{\vvec}{\boldsymbol{v}}

\newcommand{\Grad}{\boldsymbol{\nabla}}

\newcommand{\tr}{\mathrm{tr}}

\newtheorem{proposition}{Proposition}

\title{Anti-Fourier heat flux does not certify the fourth-order closure state of a rarefied cavity}
\author{Ehsan Roohi\\
Department of Mechanical and Industrial Engineering\\
University of Massachusetts Amherst, Amherst, MA 01003, USA\\
\texttt{roohie@umass.edu}}
\date{}

\begin{document}
\maketitle

\begin{abstract}
Cold-to-hot heat transfer in rarefied cavities is usually treated as a signature of Fourier-law failure.  Here it is used to ask whether a correct anti-Fourier heat-flux field certifies the flux-side fourth-order closure state.  In a two-dimensional monatomic flow, the heat-flux hierarchy observes the divergence of the composite R26-level tensor \(A_{ij}=R^{\cl}_{ij}+\Delta\delta_{ij}/3\), not the tensorial fourth-order anisotropy \(R^{\cl}_{ij}\) and scalar fourth-order excess \(\Delta\) separately.  Unlike the one-dimensional shock problem, the null space is not a single algebraic direction: it is the function space of divergence-free symmetric tensor fields, including an exactly invisible out-of-plane channel \(A_{zz}\).  DSMC data for argon lid-driven cavities show that the size of the anti-Fourier region is strongly regime dependent: it is suppressed when the lid speed is increased from \(100\) to \(200\,\mathrm{m\,s^{-1}}\), but enlarged when the Knudsen number is increased from \(0.05\) to \(0.10\).  In all cases, the anti-Fourier channel is primarily tensorial, while scalar-excess effects remain a smaller local modulation.  Hidden Airy and out-of-plane states, scaled relative to the measured RMS composite tensor, change \(R^{\cl}\) and \(\Delta\) by order-one amounts while leaving the in-plane heat-flux observable below the seed-to-seed statistical resolution, or exactly unchanged for the \(A_{zz}\) mode.  These shifted states satisfy necessary scalar Cauchy and contracted fourth-order Gram-positivity checks.  Thus anti-Fourier heat-flux agreement is a physical validation target, but it is not a certificate of full R26-level closure recovery.
\end{abstract}

\section{Introduction}
\label{sec:intro}

Heat transfer in rarefied microflows can point in directions forbidden by the Fourier law.  In lid-driven cavities, triangular cavities, backward-facing micro-steps and related confined geometries, heat lines may be directed from a colder region towards a warmer region even when the walls are isothermal.  The phenomenon itself is established.  It has been reported in kinetic simulations of rarefied cavities~\citep{JohnGuEmerson2010,JohnGuEmerson2011,MohammadzadehRoohi2012}, interpreted through nonlinear thermal-stress and stress-gradient mechanisms, and discussed using the weakly nonlinear asymptotic theory of Sone \citep{Sone2007} in \citep{mahdavi2015investigation,balaj2017regulation,MahdaviRoohi2022}.  In our earlier cavity work, DSMC calculations showed that cold-to-hot transfer depends on lid speed and rarefaction and remains consistent with the second law when entropy is evaluated from the molecular distribution \citep{MohammadzadehRoohi2012}.  In a later micro-step study, heat lines were found to deflect towards a warmer inlet region over a wide range of rarefaction, and the mechanism was explained by the competition between a Fourier-like term and higher-order velocity-curvature terms in Sone's heat-flow expansion \citep{MahdaviRoohi2022}.

The present paper asks a different question.  If a model reproduces the anti-Fourier heat-flux direction, what high-order closure information has it actually identified?  This distinction matters because moment systems do not only ask for density, velocity, temperature, stress and heat flux.  In Grad-type and regularized moment hierarchies, the heat-flux equation transports fourth-order information.  In R26, the third-order tensor \(m_{ijk}\), the fourth-order tensor \(R_{ij}\), and the scalar fourth-order excess \(\Delta\) are promoted to closure-level variables \citep{Grad1949,Struchtrup2005,StruchtrupTorrilhon2003,StruchtrupTorrilhon2007,GuEmerson2009}.  Reproducing a heat-flux vector is therefore not automatically the same as recovering the fourth-order closure state that an R26-level model would use.

This issue has a one-dimensional precursor.  In monatomic normal shocks, the heat-flux budget observes the scalar channel \(S=R^{\mathrm{cl}}_{xx}+\Delta/3\), leaving only an algebraic ambiguity in the tensorial/scalar split \citep{RoohiShock2026}.  A scalar-excess complement can close that one-dimensional split.  A two-dimensional cavity is different because the fourth-order contribution entering the in-plane heat-flux equations is not observed as \(R^{\mathrm{cl}}_{ij}\) and \(\Delta\) separately.  These two pieces enter through the composite tensor
\begin{equation}
A_{ij}=R^{\mathrm{cl}}_{ij}+\frac{1}{3}\Delta\delta_{ij},
\qquad i,j\in\{x,y\}.
\label{eq:Adef_intro}
\end{equation}
Here \(R^{\mathrm{cl}}_{ij}\) is the contracted traceless fourth-order contribution,
\(\Delta\) is the scalar fourth-order excess, and \(\delta_{ij}\) is the Kronecker delta.  Equation~\eqref{eq:Adef_intro} is therefore only a bookkeeping definition of the fourth-order channel seen by the heat-flux balance; it is not a closure assumption.  The in-plane heat-flux equations test the divergence \(\partial A_{ij}/\partial x_j\), not \(A_{ij}\) itself.  Consequently, any divergence-free symmetric addition to \(A_{ij}\) remains invisible to this observable.  The hidden space is therefore not a line, but the function space of divergence-free symmetric tensor fields.  Moreover, for a two-dimensional, three-velocity flow with \(\partial/\partial z=0\), the component \(A_{zz}\) can change the internal tensorial/scalar split while remaining absent from the in-plane heat-flux balance.  This is the new obstruction and it has no one-dimensional analogue.

The contribution of this paper is therefore not another demonstration of cold-to-hot heat transfer.  It is a flux-side R26-level observability result using anti-Fourier cavity flow as the physical setting.  We show, first, that the active anti-Fourier core in the DSMC cavity is tensorial-channel dominated, consistent with established stress-gradient and velocity-curvature interpretations.  We then show that the same heat-flux observable is compatible with order-one changes in the internal \(R^{\cl}_{ij}\)-\(\Delta\) split.  The conclusion is deliberately modest but sharp: heat-flux direction and residual agreement are not closure certificates.  A model can capture the anti-Fourier heat-flux field while leaving the fourth-order closure state underdetermined by that observable.

\section{Anti-Fourier heat transfer and the observable closure channel}
\label{sec:theory}

\subsection{Known origin of cold-to-hot transfer}

For a slightly rarefied gas, Sone's asymptotic theory writes the heat flow as a Knudsen expansion rather than as a purely Fourier response.  In the notation used in our micro-step analysis, the heat flow takes the schematic form
\begin{equation}
    \boldsymbol{Q}=\bar{\Kn}\boldsymbol{Q}_1+\bar{\Kn}^2\boldsymbol{Q}_2+\bar{\Kn}^3\boldsymbol{Q}_3,
\label{eq:sone_expansion}
\end{equation}
where \(\boldsymbol{Q}_1=0\).  The second-order term contains the Fourier-like temperature-gradient contribution
\begin{equation}
Q_{2i}=-\frac{5}{4}\gamma_2\frac{\partial \tau}{\partial x_i},
\label{eq:q2}
\end{equation}
and the third-order term contains additional nonlinear and velocity-curvature pieces, including
\begin{equation}
Q_{3i}= -\frac{5}{4}\gamma_2\frac{\partial \tau}{\partial x_i}
    -\frac{5}{4}\gamma_5\tau\frac{\partial \tau}{\partial x_i}
    +\frac{1}{2}\gamma_3\frac{\partial^2 u_i}{\partial x_j^2}+\cdots .
\label{eq:q3}
\end{equation}
In a step expansion or cavity shear layer, the higher-order velocity-curvature contribution can compete with the Fourier-like term and redirect heat lines towards a warmer region \citep{MahdaviRoohi2022}.  This explains the origin of anti-Fourier transport.  It does not answer whether the high-order closure variables responsible for that transport are identifiable.

\subsection{The two-dimensional heat-flux observable}

Let \(\cv=\vvec-\uv\) be the peculiar velocity.  The heat flux is the third-order moment
\begin{equation}
q_i=\frac{1}{2}\int |\cv|^2 c_i f\,\dd\vvec.
\label{eq:qdef}
\end{equation}
The fourth-order part of the heat-flux flux contains
\begin{equation}
B_{ij}=\int |\cv|^2 c_i c_j f\,\dd\vvec .
\label{eq:Bdef}
\end{equation}
This definition follows from the moment balance for the heat flux rather than from an algebraic closure assumption.  If the Boltzmann equation is multiplied by the heat-flux weight $\frac12 |{\bf c}|^2 c_i$ and integrated over molecular velocity, the spatial-transport term contains the flux of this third-order moment in the $(x_j)$-direction.  This introduces one additional peculiar-velocity factor $(c_j)$, giving

\[\frac{1}{2} \partial_{x_j}\int |{\bf c}|^2 c_i c_j f\,\mathrm{d}\mathbf{v}.\]

Thus, the fourth-order tensor $B_{ij}$ is the central-moment flux associated with the heat flux.  With the convention used here, the numerical factor $1/2$ is kept in the heat-flux balance, while $B_{ij}$ is defined as the fourth-order central moment itself.  The observability argument below depends on the divergence of this tensorial channel, not on the prefactor convention.

Decomposing \(c_i c_j\) into its traceless and isotropic parts gives
\begin{equation}
B_{ij}=5\rho T^2\delta_{ij}+7T\sigma_{ij}+R^{\cl}_{ij}+\frac{1}{3}\Delta\delta_{ij}.
\label{eq:Bdecomp}
\end{equation}
The lower-order terms in \eqref{eq:Bdecomp} are not the focus of the present paper.  The fourth-order closure content enters through the composite tensor \eqref{eq:Adef_intro}.  Therefore the fourth-order contribution visible to the in-plane heat-flux equations is
\begin{equation}
\mathcal{B}_i=(\Grad\cdot\Aten)_i=\partial_j A_{ij}.
\label{eq:Bobs}
\end{equation}
In a two-dimensional flow with \(\partial_z=0\) and \(u_z=0\), the observed components are
\begin{align}
\mathcal{B}_x &= \partial_x A_{xx}+\partial_y A_{xy},\label{eq:Bx}\\
\mathcal{B}_y &= \partial_x A_{xy}+\partial_y A_{yy}.\label{eq:By}
\end{align}
The component \(A_{zz}\) appears nowhere in \eqref{eq:Bx}--\eqref{eq:By}.  Yet \(A_{zz}\) contributes to the trace
\begin{equation}
\Delta = \tr \Aten = A_{xx}+A_{yy}+A_{zz},
\label{eq:traceA}
\end{equation}
and therefore changes the internal split
\begin{equation}
R^{\cl}_{ij}=A_{ij}-\frac{1}{3}\Delta\delta_{ij}.
\label{eq:RfromA}
\end{equation}
This observation already shows that the two-dimensional cavity obstruction is not the same as the one-dimensional shock obstruction.

\begin{proposition}[Two-dimensional heat-flux null space]
For a two-dimensional, three-velocity monatomic flow, the in-plane heat-flux equations observe \(\Grad\cdot\Aten\).  The in-plane observable is invariant under any perturbation \(\delta\Aten\) satisfying
\begin{equation}
\partial_j\delta A_{ij}=0,\qquad i=x,y.
\label{eq:divfree}
\end{equation}
This null space contains (i) an in-plane Airy freedom
\begin{equation}
\delta A_{xx}=\partial_{yy}\Phi,\qquad
\delta A_{xy}=-\partial_{xy}\Phi,
\qquad
\delta A_{yy}=\partial_{xx}\Phi,
\label{eq:airy}
\end{equation}
for any smooth scalar potential \(\Phi\), and (ii) an exactly invisible out-of-plane channel \(\delta A_{zz}=\eta(x,y)\), with \(\delta A_{xx}=\delta A_{xy}=\delta A_{yy}=0\).  The second freedom changes \(\Delta\) and \(R^{\cl}_{zz}\) without changing the in-plane heat-flux observable at all.
\end{proposition}

\begin{table}
\begin{center}
\small
\setlength{\tabcolsep}{5pt}
\begin{tabular}{p{0.25\textwidth}p{0.30\textwidth}p{0.34\textwidth}}
\toprule
 & One-dimensional shock & Two-dimensional cavity \\
\midrule
Observed heat-flux channel & Pointwise scalar channel \(S=A_{xx}\) & In-plane vector channel \((\Grad\cdot A)_x,(\Grad\cdot A)_y\) \\
Hidden object & Algebraic split \(S=R^{\cl}_{xx}+\Delta/3\) & Tensor field \(A_{ij}\) itself, because only its divergence is observed \\
Null space & One line: \((\delta R^{\cl}_{xx},\delta\Delta)=\eta(1,-3)\) & Function space: in-plane Airy tensors plus the exactly invisible \(A_{zz}\) channel \\
What additional data must fix & One scalar complement, e.g. \(\Delta\) & Boundary/wall information or additional moment equations for \(A\), plus out-of-plane information \\
Implication & Two-channel recovery is possible from \((S,\Delta)\) & Heat-flux direction cannot certify the internal R26-level closure state \\
\bottomrule
\end{tabular}
\caption{Comparison of the cavity and the shock test cases.  In one dimension the heat-flux budget determines a scalar fourth-order channel and leaves only an algebraic tensorial/scalar split.  In a two-dimensional, three-velocity cavity it observes only the divergence of the composite tensor \(A_{ij}=R^{\cl}_{ij}+\Delta\delta_{ij}/3\), leaving a divergence-free tensor function space and an exactly invisible out-of-plane component.}
\label{tab:schematic}
\end{center}
\end{table}

Table~\ref{tab:schematic} summarizes the resulting difference between the shock and cavity observability structures.  The 1D scalar-excess recovery \(R^{\cl}_{xx}=S-\Delta/3\) is therefore not simply waiting to be repeated in 2D.  A scalar-excess complement is still useful, but it cannot by itself determine a tensor field whose divergence is the only in-plane observed quantity.  Closing the two-dimensional null space requires information about \(A\) itself, for example through wall closure, boundary data, additional moment equations, or out-of-plane fourth-order information.

\section{DSMC cavity reference and diagnostics}
\label{sec:dsmc}

We use a two-dimensional physical-space, three-dimensional velocity DSMC reference for monatomic argon in a square lid-driven cavity of side length \(L\), with \(\Kn=\lambda_0/L\) based on the equilibrium VHS mean free path at the wall temperature.  All walls are diffuse and isothermal at \(T_w=300\,\mathrm{K}\); the top wall moves with speed \(U_w=100\,\mathrm{m\,s^{-1}}\).  The base case has \(\Kn=0.05\), a \(160\times160\) sampling mesh, fixed collision subcells, an upstream-equivalent particle loading of 128 particles per cell, and eight independent random seeds.  The argon VHS parameters are \(m=6.6335\times10^{-26}\,\mathrm{kg}\), \(d_{\rm ref}=4.17\times10^{-10}\,\mathrm{m}\) at \(T_{\rm ref}=273\,\mathrm{K}\), and viscosity index \(\omega=0.81\) \citep{Bird1994}.  Collisions are sampled with Bird's no-time-counter procedure.  The time step is kept below the cell-crossing and mean-collision-time scales; after transient removal, samples are accumulated at a fixed stride, giving 8501 samples per realisation and an average collision count of \(9.32\times10^4\) per cell in the ensemble used below.

The sampled fields are \(\rho\), \(\uv\), \(T\), \(\qv\), \(\sigma_{ij}\), \(R^{\cl}_{ij}\), \(\Delta\), and the composite tensor \(A_{ij}\).  The present production runs do not sample \(m_{ijk}\).  This does not affect the observability result in \S\ref{sec:theory}, because \(m_{ijk}\) does not enter the fourth-order flux of the heat-flux equation; it enters the stress and R26 evolution equations.  A full R26 audit or an R26 cavity solve would require \(m_{ijk}\), but the present claim is narrower: it concerns which part of the measured R26-level closure state is visible to the heat-flux channel.  Additional R26 evolution equations and wall boundary conditions can constrain the closure state; the point here is that agreement of the heat-flux observable, including its anti-Fourier direction, cannot by itself certify that state.

We identify anti-Fourier regions using the cosine indicator
\begin{equation}
I_{AF}=\frac{\qv\cdot\Grad T}{|\qv|\,|\Grad T|}.
\label{eq:IAF}
\end{equation}
Positive $I_{AF}$ means that heat flux points towards increasing temperature.  Since the anti-Fourier diagnostic is itself directional, the relevant fourth-order contribution is the component of $(\nabla\cdot A)$ along the locally observed heat-flux direction.  We therefore use $(\boldsymbol q/|\boldsymbol q|)$ as the local projection direction and resolve the tensorial and scalar parts of this fourth-order channel as
\begin{equation}
P_R=\frac{\boldsymbol q\cdot(\nabla\cdot R^{\mathrm{cl}})}{|\boldsymbol q|},
\label{eq:PR}
\end{equation}
\begin{equation}
P_\Delta=\frac{\boldsymbol q\cdot(\nabla\Delta/3)}{|\boldsymbol q|}.
\label{eq:PD}
\end{equation}
These projections measure the signed tensorial and scalar contributions to the heat-flux-parallel fourth-order channel, rather than introducing an independent closure criterion.

Unless otherwise stated, fields are smoothed with a seven-point centred window before differentiating.  The active region is defined by simultaneous lower bounds on \(|\qv|\) and \(|\Grad T|\), with a threshold 0.05 relative to their domain maxima.  The same thresholding and smoothing rule is used for all reported cases.

\section{The anti-Fourier core is tensorial-channel dominated}
\label{sec:driver}

\begin{figure}
  \centerline{\includegraphics[width=1.0\textwidth]{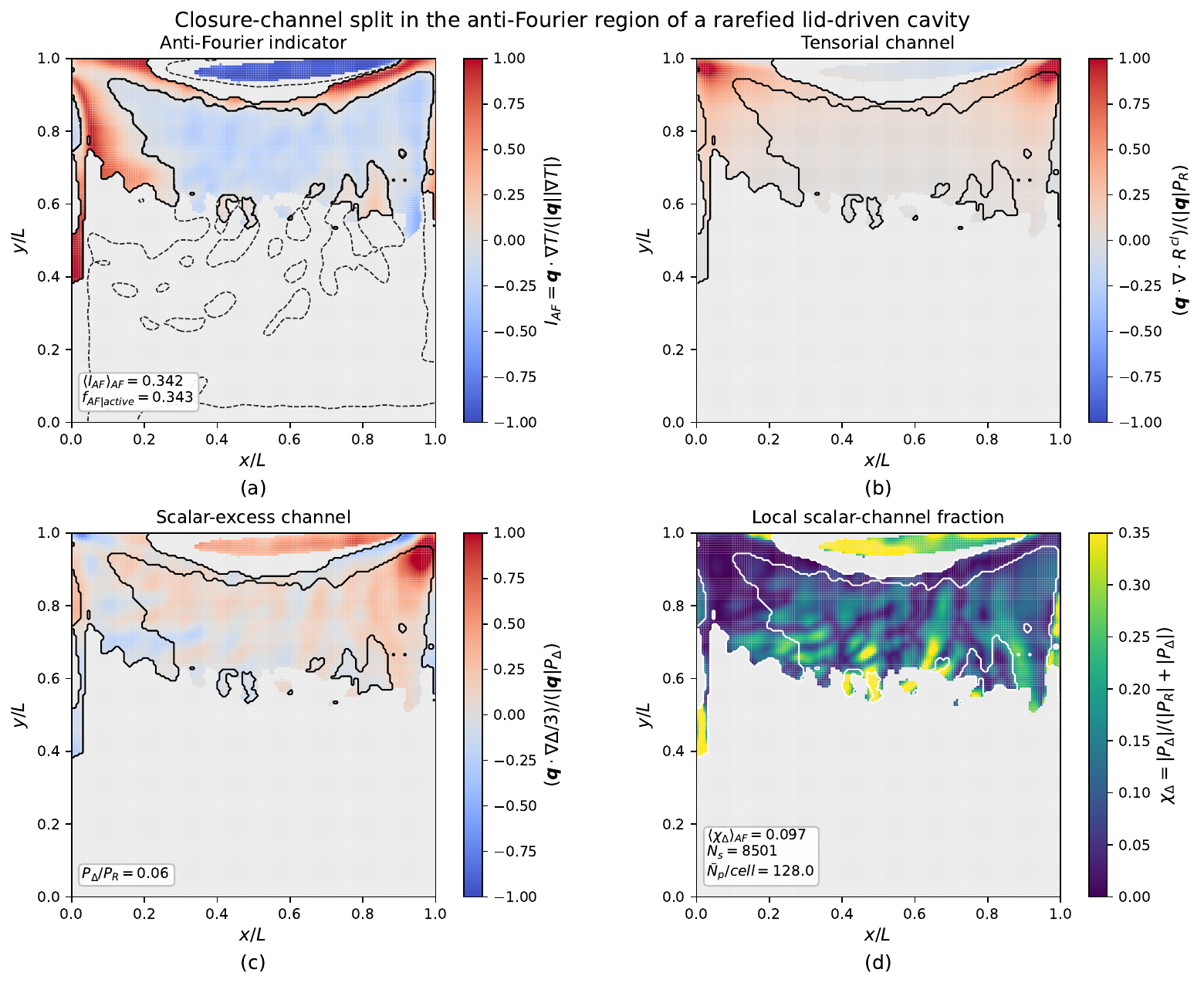}}
  \caption{Closure-channel split in the base DSMC cavity, \(\Kn=0.05\), \(U_w=100\,\mathrm{m\,s^{-1}}\).  The panels show (a) \(I_{AF}\), (b) \(\qv\cdot(\Grad\cdot R^{\cl})/|\qv|\), (c) \(\qv\cdot\Grad\Delta/(3|\qv|)\), and (d) \(\chi_\Delta\).  Panels (b,c) use separate normalisations; \(P_\Delta/P_R=0.063\).  The active anti-Fourier core is tensorial-channel dominated.}
  \label{fig:cavity_split}
\end{figure}

Figure~\ref{fig:cavity_split} shows the closure-channel split in the base case.  The anti-Fourier region occupies the upper portion of the cavity, especially near the driven lid and sidewall regions.  Within the active region, 34.3\% of cells are anti-Fourier and the mean anti-Fourier indicator in that subset is \(\langle I_{AF}\rangle_{AF}=0.342\).  The tensorial projection \(P_R\) is coherent and positive in the same region.  The scalar projection \(P_\Delta\) has visible local structure but is much smaller: its measured scale is only \(P_\Delta/P_R=0.063\), and the mean scalar fraction is \(\langle\chi_\Delta\rangle_{AF}=0.097\).

\begin{figure}
  \centerline{\includegraphics[width=0.72\textwidth]{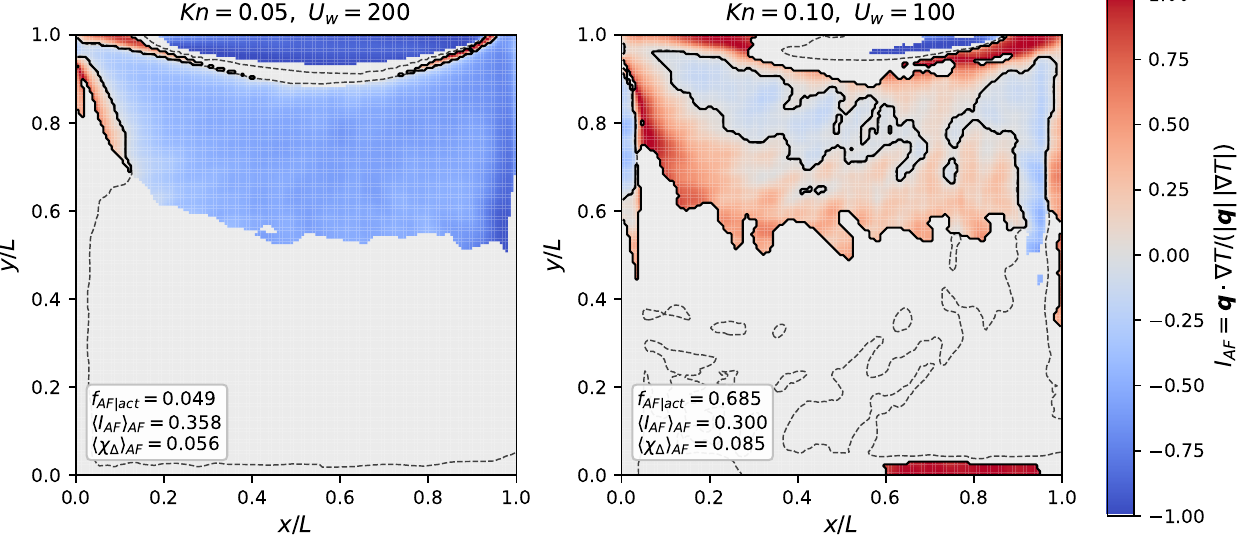}}
  \caption{Robustness of the anti-Fourier region across lid speed and rarefaction.  The base case is shown in figure~\ref{fig:cavity_split}; this figure shows only the two non-overlapping robustness cases.  The colour field is the anti-Fourier indicator \(I_{AF}=\qv\cdot\Grad T/(|\qv|\,|\Grad T|)\), shown in the active region used for the closure-channel diagnostics.  The solid contour marks the active anti-Fourier set.  Increasing the lid speed suppresses the anti-Fourier area, whereas increasing \(\Kn\) enlarges it.  The scalar-channel fraction remains small in all cases, so the physical extent of the anti-Fourier region changes more strongly than the tensorial dominance of the observed channel.}
  \label{fig:robustness}
\end{figure}

The behaviour is robust to changes in lid speed and rarefaction, but the physical size of the anti-Fourier region is not.  The reported fractions use the same active-threshold definition in all cases and are used as regime descriptors rather than as a new universal measure.  Since the base case has already been shown in figure~\ref{fig:cavity_split}, figure~\ref{fig:robustness} shows only the two additional robustness cases: a higher-speed lid and a more rarefied cavity.  Increasing the lid speed from \(U_w=100\) to \(200\,\mathrm{m\,s^{-1}}\) suppresses the anti-Fourier area: the active anti-Fourier fraction drops from 0.343 in the base case to 0.049.  By contrast, increasing the Knudsen number from \(0.05\) to \(0.10\) greatly expands the anti-Fourier region, giving an active anti-Fourier fraction of 0.685.  The scalar-channel fraction remains modest in all cases, \(\langle\chi_\Delta\rangle_{AF}\leq 0.097\), so the anti-Fourier direction remains primarily tensorial even as its spatial support changes substantially.

\begin{table}
\begin{center}
\small
\setlength{\tabcolsep}{4pt}
\begin{tabular}{lcccccccc}
\toprule
case & \(f_{AF|act}\) & \(\langle I_{AF}\rangle_{AF}\) & \(\langle\chi_\Delta\rangle_{AF}\) & \(\sigma_{seed}/\|P_A\|\) & \(\delta R_A\) & \(\delta\Delta_A\) & \(\delta R_{zz}\) & \(\delta\Delta_{zz}\) \\
\midrule
\(\Kn=0.05,\ U_w=100\) & 0.343 & 0.342 & 0.097 & 0.071 & 0.63 & 0.35 & 0.59 & 0.40 \\
\(\Kn=0.05,\ U_w=200\) & 0.049 & 0.358 & 0.056 & 0.030 & 0.47 & 0.60 & 0.46 & 0.62 \\
\(\Kn=0.10,\ U_w=100\) & 0.685 & 0.300 & 0.085 & 0.062 & 0.52 & 0.46 & 0.51 & 0.48 \\
\bottomrule
\end{tabular}
\caption{Robustness of anti-Fourier observability and hidden closure freedom.  The null-space changes are reported for \(\alpha=0.5\), for which the scalar Cauchy bound and contracted fourth-order Gram-positivity checks have zero violations in all cases.  Here \(\delta R_A,\delta\Delta_A\) denote the relative changes induced by the in-plane Airy mode, and \(\delta R_{zz},\delta\Delta_{zz}\) those induced by the exactly invisible out-of-plane mode.}
\label{tab:robustness}
\end{center}
\end{table}

\section{Divergence-free closure states with identical heat-flux observable}
\label{sec:nullspace}

We now test the two-dimensional null space directly on the DSMC field.  The perturbations are scaled relative to the RMS size of the composite tensor \(A\) in the active region; thus \(\alpha=0.5\) denotes a perturbation whose RMS tensor magnitude is one half of the RMS magnitude of the measured composite tensor before differentiation.  For each shifted state, we reconstruct
\begin{equation}
\Delta^*=\tr \Aten^*,\qquad R^{\cl *}_{ij}=A^*_{ij}-\frac{1}{3}\Delta^*\delta_{ij},
\label{eq:shift_recover}
\end{equation}
and monitor both the observed projected channel
\begin{equation}
P_A=\frac{\qv\cdot(\Grad\cdot\Aten)}{|\qv|}
\label{eq:PA}
\end{equation}
and the internal changes in \(R^{\cl}\) and \(\Delta\).  The eight independent seeds provide the RMS seed-to-seed scatter of \(P_A\) in the active region, \(3.56\times10^{39}\), while the RMS mean observed channel is \(5.03\times10^{40}\).  We also check the necessary scalar Cauchy margin \(C_\Delta=\Delta^*+6n\theta^2>0\).

Table~\ref{tab:nullspace} reports two classes of shifts after adding admissibility checks.  The first check is the Cauchy--Schwarz consequence
\begin{equation}
    \left(\int |\cv|^2 f\,\dd\vvec\right)^2
    \leq \left(\int f\,\dd\vvec\right)
          \left(\int |\cv|^4 f\,\dd\vvec\right),
\end{equation}
which gives the scalar lower bound \(\Delta^*\geq -6n\theta^2\).  The second check is tensorial: for any vector \(\boldsymbol a\),
\begin{equation}
    a_i a_j B^*_{ij}=\int |\cv|^2(\boldsymbol a\cdot\cv)^2 f^*\,\dd\vvec \geq 0,
\end{equation}
so the contracted fourth-order matrix
\begin{equation}
    B^*_{ij}=5n\theta^2\delta_{ij}+7\theta\sigma_{ij}+A^*_{ij}
    \label{eq:Bpsd}
\end{equation}
must be positive semidefinite.  Here \(n\) and \(\theta\) denote the number-density and thermal-temperature variables used in the DSMC sampling; this is the same contracted fourth-order tensor as \(5\rho T^2\delta_{ij}+7T\sigma_{ij}+A_{ij}\) in the mass-normalised notation of \eqref{eq:Bdecomp}, up to the constant molecular-mass scaling.  These are necessary, not sufficient, realizability conditions.  They are included to ensure that the hidden-state amplitudes are not supported only by an arbitrary algebraic perturbation.

The Airy mode changes the in-plane tensor components according to \eqref{eq:airy}.  Its observable change is not exactly zero on the discrete grid, but it is far below the seed-to-seed envelope: for \(\alpha=0.5\), \(\|P_A^*-P_A\|/\|P_A\|=1.46\times10^{-13}\), the correlation with the original channel is 1.0, and the change is only \(2.06\times10^{-12}\) of the seed standard deviation.  Nevertheless \(R^{\cl}\) changes by 63\% and \(\Delta\) by 35\%.  The Cauchy margin and the minimum eigenvalue of \(B^*\) remain positive.  The out-of-plane \(A_{zz}\) mode is even cleaner: it changes neither \(A_{xx}\), \(A_{xy}\) nor \(A_{yy}\), and therefore the in-plane heat-flux observable is exactly unchanged.  At \(\alpha=0.5\), this invisible channel changes \(R^{\cl}\) by 59\% and \(\Delta\) by 40\% while satisfying the same necessary checks.

\begin{table}
\begin{center}
\begin{tabular}{ccccccc}
\toprule
mode & \(\alpha\) & \(\|\delta P_A\|/\sigma_{seed}\) & \(\|\delta R^{\cl}\|/\|R^{\cl}\|\) & \(\|\delta\Delta\|/\|\Delta\|\) & \(\min C_\Delta\) & \(\min\lambda_B\) \\
\midrule
Airy in-plane & 0.25 & \(1.03\times10^{-12}\) & 0.313 & 0.175 & 4.5660 & 3.2633 \\
Airy in-plane & 0.50 & \(2.06\times10^{-12}\) & 0.627 & 0.350 & 4.5661 & 3.2633 \\
Airy in-plane & 1.00 & \(4.12\times10^{-12}\) & 1.253 & 0.700 & 4.5663 & 3.2634 \\
\midrule
Invisible \(A_{zz}\) & 0.25 & 0 & 0.297 & 0.199 & 4.5515 & 3.2633 \\
Invisible \(A_{zz}\) & 0.50 & 0 & 0.594 & 0.397 & 4.5370 & 3.2633 \\
Invisible \(A_{zz}\) & 1.00 & 0 & 1.189 & 0.794 & 4.5082 & 3.2633 \\
\bottomrule
\end{tabular}
\caption{Necessary-admissibility checks for hidden closure states in the base cavity.  The seed-to-seed RMS scatter of the observed projected channel is \(\sigma_{seed}=3.56\times10^{39}\), about 7.1\% of the RMS mean channel.  Here \(C_\Delta=\Delta^*+6n\theta^2\) is the scalar Cauchy margin and \(\lambda_B\) is the minimum eigenvalue of the contracted fourth-order matrix \(B^*\) in \eqref{eq:Bpsd}; the last two columns, reported in units of \(10^{35}\), are recomputed on the shifted states.  The Airy mode remains far below the statistical resolution of the observed channel, while the \(A_{zz}\) mode is exactly invisible.  All displayed rows satisfy \(C_\Delta>0\) and \(B^*\succeq0\) throughout the active region; a sweep up to \(\alpha=5\) also gives zero violations for these necessary checks, but we report the conservative \(\alpha\leq1\) range here.}
\label{tab:nullspace}
\end{center}
\end{table}

This is the central result.  The cavity heat-flux field, including its anti-Fourier direction, is compatible with strongly different fourth-order closure states that pass necessary scalar and contracted-tensor realizability checks.  The result does not depend on claiming that \(\Delta\) drives the heat flux.  Even in a tensorial-dominated anti-Fourier core, \(\Grad\cdot A\) does not identify \(A\), and \(A\) does not uniquely certify the internal \(R^{\cl}\)-\(\Delta\) split without further information.  A full sufficient realizability characterization for the shifted moment sets is beyond the scope of this paper; the qualitative non-certification is exact, while the numerical amplitudes demonstrate that the hidden freedom is not confined to infinitesimal changes under standard necessary checks.

\section{Implications for R13, R26 and kinetic-model validation}
\label{sec:discussion}

The present paper should be read as an R26-level observability diagnostic, not as a full R26 cavity calculation.  R13 does not contain \(R_{ij}\) and \(\Delta\) as independent variables; in R13 they are slaved through regularized closures.  R26 promotes these quantities to independent closure-level variables.  Therefore the present question is intrinsically R26-level, even though the reference data are obtained from DSMC rather than from an R26 solver.

A full R26 calculation would of course include evolution equations and wall boundary conditions for the higher moments, and those equations may select one admissible closure state.  Our claim is narrower: a validation practice based on the heat-flux vector, its anti-Fourier direction, or the in-plane heat-flux balance cannot by itself certify that selected state.  This is an observability statement about the flux-side channel, not a claim that the full R26 system is unsolvable or non-unique.

What would close the two-dimensional null space?  The one-dimensional answer, a scalar-excess budget, is necessary but not sufficient.  The in-plane Airy freedom requires information that constrains the tensor field itself, such as boundary/wall closure on \(A\), additional moment equations, or direct kinetic information.  This is consistent with recent work showing that higher-moment boundary conditions are part of the physical closure problem: Onsager-stable R13 and LR26 boundary formulations impose wall information compatible with thermodynamic admissibility and reciprocity \citep{TorrilhonStruchtrup2008,CaiTorrilhonYang2024,RanaGuptaSprittlesTorrilhon2021}.  The present DSMC construction does not contradict such closures.  It shows that the heat-flux vector alone does not supply the wall or moment information needed to determine the hidden in-plane tensor field or the out-of-plane channel \(A_{zz}\).

\section{Conclusions}
\label{sec:conclusion}

The anti-Fourier heat flux of a rarefied cavity is a useful physical validation target, but it is not a full closure certificate.  The heat-flux hierarchy observes the divergence of the composite fourth-order tensor \(A_{ij}=R^{\cl}_{ij}+\Delta\delta_{ij}/3\).  In a two-dimensional cavity this leaves a divergence-free tensor null space and an exactly invisible out-of-plane channel, both absent from the one-dimensional shock.  DSMC data for an argon lid-driven cavity show that the strong anti-Fourier core is mainly tensorial-channel dominated, while the scalar-excess channel is a smaller local modulation.  More importantly, hidden Airy and out-of-plane states can leave the heat-flux observable below statistical resolution, or identically unchanged, while altering \(R^{\cl}\) and \(\Delta\) by order-one amounts under necessary scalar and contracted fourth-order Gram-positivity checks.

The paper therefore reframes anti-Fourier heat transfer as a test of closure observability.  Reproducing the heat-flux direction demonstrates that a model captures an important physical signature; it does not demonstrate that the model has recovered the full R26-level fourth-order closure state.  The additional lid-speed and rarefaction tests show that the size of the anti-Fourier region can shrink or expand substantially while the hidden-space obstruction persists.  A robust high-order closure for multidimensional rarefied heat transfer must identify the observed composite channel and supply additional information to fix the hidden tensorial/scalar split.

\section*{Acknowledgements}
The author thanks the rarefied-gas-dynamics community whose studies of cavity heat transfer, moment closures and kinetic particle methods motivated the present work.

\section*{Funding}
This work received no specific grant from any funding agency, commercial or not-for-profit sector.

\section*{Declaration of interests}
The author reports no conflict of interest.

\section*{Data availability statement}
The DSMC post-processing data and plotting scripts can be made available upon reasonable request.

\section*{Declaration of AI-assisted technologies}
The author used ChatGPT (OpenAI, GPT-5.5 Pro, accessed May 2026) for language
editing and manuscript polishing. The author reviewed, edited and verified all
scientific content, derivations, numerical values and references, and takes full
responsibility for the manuscript.

\end{document}